\documentclass[11pt, oneside,reqno]{amsart}

\usepackage[top=2.5cm, bottom=2.5cm, left=2.5cm, right=2.5cm]{geometry}

\usepackage[english]{babel}

\usepackage[T1]{fontenc}

\usepackage{textcomp}

\usepackage[utf8]{inputenc}

\usepackage{amsfonts, amsmath, amssymb, mathtools, amsthm}
\usepackage[makeroom]{cancel}
\usepackage[none]{hyphenat}
\usepackage{bigints}
\usepackage{xcolor}

\DeclarePairedDelimiter{\abs}{\lvert}{\rvert}

\usepackage{siunitx}
\usepackage{fancyhdr}
\usepackage{graphicx}
\usepackage{verbatim}
\usepackage{xcolor}
\usepackage{pgf, tikz, tkz-base}
\usepackage{mathrsfs}
\usepackage{tkz-euclide}
\usepackage{xfp}

\usetikzlibrary{shapes, calc, shapes, arrows, babel}
\usepackage[margin=16pt,font=small,labelfont=bf]{caption} 

\usepackage[maxbibnames=99, sorting=nyt]{biblatex}
\usepackage{parskip}

\usepackage[colorlinks=true]{hyperref}

\begin{document}

\title
{Dumbbell dynamics: a didactical approach}

\author{
Benedetto Scoppola$^{1}$ \and
Matteo Veglianti$^{2}$}

\date{}

\maketitle

\begin{center}
{\footnotesize
\vspace{0.3cm}$^{1}$  Dipartimento di Matematica,\\
Universit\`a di Roma
``Tor Vergata''\\
Via della Ricerca Scientifica - 00133 Roma, Italy\\
\texttt{scoppola@mat.uniroma2.it}\\

\vspace{0.3cm} $^{2}$ Dipartimento di  Fisica,\\ Universit\`a di Roma
``Tor Vergata''\\
Via della Ricerca Scientifica - 00133 Roma, Italy\\
\texttt{matteoveglianti@gmail.com}\\ 
}

\end{center}

\begin{abstract}
In this paper we propose a simplified model to describe the dissipative effects of tides. We assume a spherical Earth with a dissipative coupling with a mechanical dumbbell. The latter has a mass much smaller than the Earth's, and it models the presence of the tidal bulges.
Using properly the scale analysis, we will show that some of the consequences of tidal dissipation are the circularization and the enlargement of orbit of the Moon and the slowing down of the Earth's rotation. We will also see that tidal dissipation plays a fundamental role for the establishment of a regime of spin-orbit resonance in the celestial systems. The mathematical tools used make our treatment appropriate for senior high school students or college students.
\end{abstract}

\section{Introduction}
All textbooks in introductory astronomy and many in physics and mechanics mention the existence of oceanic tides as an interesting manifestation of universal gravitation: indeed many teachers are interested in this topic. As argued in \cite{1}, the most important aspects of the origin and properties of tides are often treated inaccurately or even erroneously. Much of the confusion over generating tides is related to the roles of the orbital motion of the Moon and earth about their common center of mass and of the Earth's axial rotation. In discussing the physics behind this phenomenon, authors usually explain (more or less successfully) why two tidal swells appear on the opposite sides of the globe. However, it is difficult to find a plausible explanation of the physical mechanism responsible for the phase shift between the zenith of the moon and the moment of high tide, which at some places approaches 90°. Misunderstandings also occur in discussions about the role of tidal friction in the retardation of axial rotations and in the evolution of orbital motions of the gravitationally coupled celestial bodies.\\
While the conservative aspects of the tides are masterfully treated using elementary tools in \cite{1}, \cite{2} and \cite{3}, the dissipative ones are only qualitatively described. The scientific, non-pedagogical, works on tidal dissipation are divided into two large areas according to the desired target: rheological aspects (see \cite{4} and \cite{5})  or dynamic aspects (see \cite{6} and \cite{9}). Our aim is therefore to propose a way to quantitatively deal with the dissipative aspects of the tides and their consequences using high school mathematics. To this end, we will use the "dumbbell model" developed in \cite{7}, that consists in describing the planet in terms of a point $P$ of mass $M-\mu$ and a mechanical dumbbell centered in $P$, i.e., a system of two points, each of which has mass $\mu/2$, constrained to be at fixed mutual distance $2r$, having $P$ as center of mass.\\
The idea of a dumbbell model is not original, in fact is developed in \cite{9} and in many works (see, for instance \cite{8}) where however it is used for other purposes.\\
This model is useful to compute, using an elementary force's approach, the torque acting on the Earth’s ocean bulges due to the Moon and the torque acting on the Moon due to the Earth’s ocean bulges. We perform the detailed computation in section \ref{section_torque}.
In section \ref{section_evolution} we present a way to describe the evolution of the system imagining that the variation of the parameters does not occur in a continuous way, but rather discretely. This allows us to avoid a treatment through differential equations and replace them with finite difference equations.\\
In this way, it will be possible to show how the tidal dissipation is responsible for the circularization and the enlargement of the lunar orbit and the slowing down of the earth's rotation. We will see that the first two events occur on very different time scales: the circularization of the lunar orbit first is much faster than the enlargement. Thus the orbit will become circular in shorter times than those of enlargement, and this is what we see in many planet-satellite systems of our galaxy, particularly in the Earth-Moon system. Finally, we will also see that tidal dissipation plays a fundamental role for the establishment of a regime of spin-orbit resonance in the celestial systems.\\
The mathematical tools used make our discussion appropriate for senior high school students or college students. We believe that an appropriate and non-sterile use of mathematics is useful to understand its functionality and to make children passionate about studying this discipline. Furthermore, this subject is very suitable for teaching and learning the scale analysis, that is a very powerful tool to simplify complex equations by neglecting the suitable small terms.

\section{Tidal torque}
\label{section_torque}
\begin{figure}[h]
	\includegraphics[scale=0.4]{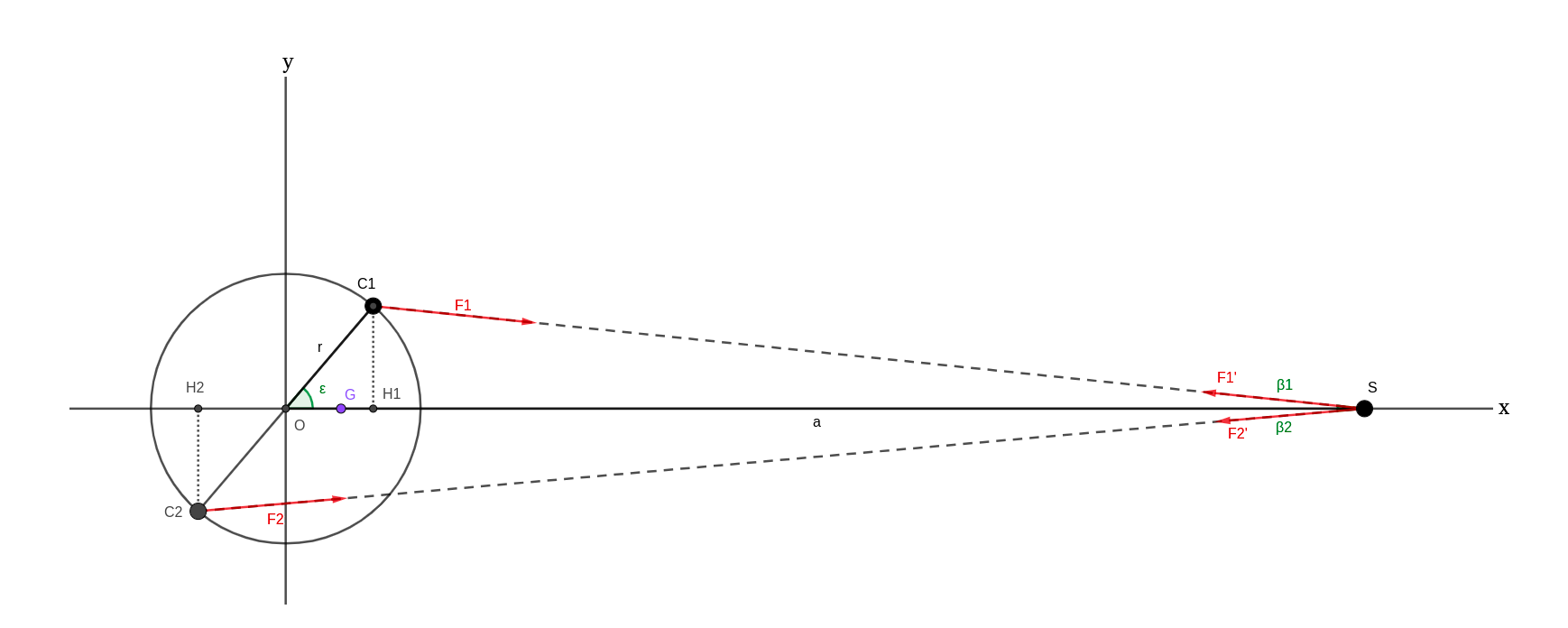}
	\caption{Geometry of the system.}
	\label{torque}
\end{figure}

In this section we want to show how the dumbbell model is useful to compute the torque acting on the Earth's ocean bulges due to the Moon. 
The ocean bulges are modeled by the aforementioned dumbbell: a pair of massive point each of which with a mass of $\mu /2$ placed at the ends of a segment of length $2r$, that we can assume, for simplicity, equal to the diameter $2R$ of the Earth.\footnote{With a detailed computation using suitable integrals, one can show that $r=\frac{3}{4}R$.}\\
In figure \ref{torque} we imagine the Earth as a sphere with radius $R$ centered at the origin $O$ of the reference frame and the Moon as a massive point, indicated with $S$, with mass $m$, at a distance $a$ from the center of the Earth. Let the $x$-axis be the line joining $O$ with $S$ and the $y$-axis perpendicular to it. Moreover, for simplicity, we consider the rotation of the Earth perpendicular to the Moon's orbital plane.\\
On the Earth's surface there are the two massive points $C_1$ and $C_2$, that represent the ocean bulge's center of gravity. Finally, in general, we imagine that the dumbbell is inclined by an angle $\varepsilon$ with respect to the $x$-axis.\\
 Let $\vec{F_1}$ and $\vec{F_2}$ be the attractive forces acting on $C_1$ and $C_2$ due to the Moon and let $\vec{F_1'}$ and $\vec{F_2'}$ the attractive forces acting on $S$ due to $C_1$ and $C_2$ respectively. Clearly $\vec{F_1'}=-\vec{F_1}$ and $\vec{F_2'}=-\vec{F_2}$.\\
Moreover, on $S$ also acts the attractive force $\vec{F}$, lying on the $x$-axis, due to the rest of the Earth, deprived of the two masses of water.\\

Let $G$ be the universal gravitational constant and let $\beta_1$ and $\beta_2$ be the two angles $O{\hat S}C_1$ and $O{\hat S}C_2$ respectively; we have:
\begin{equation}
	\label{F1}
	F_1=-F_1'=-\frac{Gm\frac{\mu}{2}}{SC_1^2}(-\hat{x} \cos \beta_1 + \hat{y} \sin \beta_1),
\end{equation}
\begin{equation}
	\label{F2}
	F_2=-F_2'=-\frac{Gm\frac{\mu}{2}}{SC_2^2}(-\hat{x} \cos \beta_2 - \hat{y} \sin \beta_2),
\end{equation}
\begin{equation}
	\label{F}
	F=-\frac{Gm(M- \mu)}{a^2}\hat{x}.
\end{equation}

Moreover, from the geometry of the system we have that $\frac{r}{a}$ is a dimensionless small parameter, so we can expand the forces up to the second order in $\frac{r}{a}$.\\
To this end, we need the following geometric relation:
\begin{equation}
	\label{geometry}
	\begin{matrix} 
		SC_1^2=(a-r \cos \varepsilon)^2+(r \sin \varepsilon)^2 = a^2 \biggl( 1 - 2\frac{r}{a} \cos \varepsilon + \frac{r^2}{a^2} \biggr) \\
		SC_2^2=(a+r \cos \varepsilon)^2+(r \sin \varepsilon)^2 = a^2 \biggl( 1 + 2\frac{r}{a} \cos \varepsilon + \frac{r^2}{a^2} \biggr) \\
		\beta_1=\frac{r \sin \varepsilon}{a - r \cos \varepsilon} = \frac{r}{a} \sin \varepsilon \biggl( 1 + \frac{r}{a} \cos \varepsilon \biggr)\\ 
		\beta_2=\frac{r \sin \varepsilon}{a + r \cos \varepsilon} = \frac{r}{a} \sin \varepsilon \biggl( 1 - \frac{r}{a} \cos \varepsilon \biggr) 
	\end{matrix}
\end{equation}

From (\ref{F1}) we have:
\begin{equation*}
	F_1=-F_1'=-\frac{Gm\frac{\mu}{2}}{a^2} \biggl( 1 + 2\frac{r}{a} \cos \varepsilon - \frac{r^2}{a^2} + 4\frac{r^2}{a^2} \cos^2 \varepsilon \biggr) \biggl[-\hat{x} \biggl( 1 - \frac{r^2}{2a^2} \sin^2 \varepsilon \biggr) + \hat{y} \biggl( \frac{r}{a} \sin \varepsilon \biggl( 1 + \frac{r}{a} \cos \varepsilon \biggr) \biggr) \biggr]=
\end{equation*}
\begin{equation}
	\label{F1compute}
	=-\frac{Gm\frac{\mu}{2}}{a^2} \biggl[-\hat{x} \biggl( 1 + 2\frac{r}{a} \cos \varepsilon - \frac{r^2}{a^2}\biggl( \frac{3}{2} - \frac{9}{2} \cos^2 \varepsilon \biggr) \biggr) + \hat{y} \biggl(\frac{r}{a} \sin \varepsilon \biggl( 1 + 3\frac{r}{a} \cos \varepsilon \biggr) \biggr) \biggr].
\end{equation}

Similarly, from (\ref{F2}) we have:
\begin{equation}
	\label{F2compute}
	F_2=-F_2'=-\frac{Gm\frac{\mu}{2}}{a^2} \biggl[-\hat{x} \biggl( 1 - 2\frac{r}{a} \cos \varepsilon - \frac{r^2}{a^2}\biggl( \frac{3}{2} - \frac{9}{2} \cos^2 \varepsilon \biggr) \biggr) - \hat{y} \biggl(\frac{r}{a} \sin \varepsilon \biggl( 1 - 3\frac{r}{a} \cos \varepsilon \biggr) \biggr) \biggr].
\end{equation}

We are now ready to calculate the torque acting on the dumbbell due to the Moon and the torque acting on the Moon due to the dumbbell. We will show that the two torques are exactly opposite, as it is expected from the conservation of angular momentum for isolated systems.\\
For simplicity we impose that the Moon moves in a circular orbit around the point G, the center of gravity of the Earth-Moon system. This is equivalent to stating that the sum of the components along the $x$-axis of the forces acting on the Moon is the centripetal force.
Let $\omega$ be the angular velocity of the revolution of the Moon and $r_S=SG$ the radius of the circular orbit around the point $G$, so we have:
\begin{equation*}
	m\omega^2 r_S = F_x+F'_{1x}+F'_{2x} =
\end{equation*}
\begin{equation*}
	= \frac{Gm(M- \mu)}{a^2}+\frac{Gm\frac{\mu}{2}}{a^2} \biggl( 1 + 2\frac{r}{a} \cos \varepsilon - \frac{r^2}{a^2}\biggl( \frac{3}{2} - \frac{9}{2} \cos^2 \varepsilon \biggr) \biggr) + \frac{Gm\frac{\mu}{2}}{a^2} \biggl( 1 - 2\frac{r}{a} \cos \varepsilon - \frac{r^2}{a^2}	\biggl(	\frac{3}{2} - \frac{9}{2} \cos^2 \varepsilon \biggr) \biggr)=
\end{equation*}
\begin{equation*}
	= \frac{GmM}{a^2} - \frac{Gm\mu}{a^2} \frac{3}{2} \frac{r^2}{a^2} (1-3\cos^2 \varepsilon)
\end{equation*}
Hence:
\begin{equation}
	\label{rs}
	r_S = \frac{GM}{\omega^2 a^2} \biggl(1 +\frac{3}{2} \frac{\mu}{M} \frac{r^2}{a^2} (3\cos^2 \varepsilon -1) \biggr).
\end{equation}
Notice that, in the case of two point masses $m$ and $M$ placed at a distance $a$, it turns out that $m$ makes a circular orbit around the common center of gravity with a radius $r_S=\frac{GM}{\omega^2 a^2}$.\\
Since in (\ref{rs}) both $\frac{\mu}{M}$ and $\frac{r^2}{a^2}$ are small quantities, the correction to the Moon's orbital radius due to the dumbbell is completely negligible.\\

We can now compute both the torque acting on the dumbbell due to the Moon and the torque acting on the Moon due to the dumbbell up to the smallest order in $\frac{r}{a}$.\\
Let's start with the latter. The magnitude of the torque acting on the Moon is $\Gamma_M=a\abs{\vec{F_M}}$, with $\vec{F_M}$ is the sum of the forces acting on the Moon. Thanks to (\ref{rs}), $\vec{F_M}$ is parallel to the $y$-axis, and the magnitude of the torque is:
\begin{equation*}
	\Gamma_M = a (F'_{1y}-F'_{2y}) = \frac{Gm\frac{\mu}{2}}{a^2} \biggr[\frac{r}{a} \sin \varepsilon \biggl( 1 + 3\frac{r}{a} \cos \varepsilon \biggr)-\frac{r}{a} \sin \varepsilon \biggl( 1 - 3\frac{r}{a} \cos \varepsilon \biggr) \biggl]=
\end{equation*}
\begin{equation}
	\label{gammam}
	=3\frac{r^2}{a^3}Gm \mu \sin\varepsilon \cos\varepsilon.
\end{equation} 
 
On the other hand, the magnitude of the torque acting on the dumbbell due to the Moon is:
\begin{equation*}
	\Gamma_D = r \abs{\vec{F_1}} \sin(\varepsilon+\beta_1)-r \abs{\vec{F_2}} \sin(\varepsilon - \beta_2)=
\end{equation*}
\begin{equation*}
	= \frac{rGm\frac{\mu}{2}}{a^2} \biggr( 1 + 2\frac{r}{a} \cos \varepsilon -\frac{r^2}{a^2} + 4\frac{r^2}{a^2} \cos^2 \varepsilon  \biggl) \sin(\varepsilon+\beta_1) - \frac{rGm\frac{\mu}{2}}	{a^2} \biggr( 1 - 2\frac{r}{a} \cos \varepsilon - \frac{r^2}{a^2} + 4\frac{r^2}{a^2} \cos^2 			\varepsilon  \biggl) \sin(\alpha-\beta_2)=
\end{equation*}
\begin{equation*}
	= \frac{rGm\frac{\mu}{2}}{a^2} \biggr( 1 + 2\frac{r}{a} \cos \varepsilon  \biggl) (\sin\varepsilon\cos\beta_1+\cos\varepsilon\sin\beta_1) - \frac{rGm\frac{\mu}{2}}{a^2} \biggr( 1 	- 2\frac{r}{a} \cos \varepsilon \biggr) (\sin\varepsilon\cos\beta_2-\cos\varepsilon\sin\beta_2)=
\end{equation*}
\begin{equation*}
	= \frac{rGm\frac{\mu}{2}}{a^2} \biggr( 1 + 2\frac{r}{a} \cos \varepsilon  \biggl) (\sin\varepsilon+\frac{r}{a}\cos\varepsilon\sin\varepsilon) - \frac{rGm\frac{\mu}{2}}{a^2} \biggr( 1 - 2\frac{r}{a} \cos \varepsilon \biggr) (\sin\varepsilon+\frac{r}{a} \cos\varepsilon\sin\varepsilon)=
\end{equation*}
\begin{equation}
	\label{gammad}
	=3\frac{r^2}{a^3}Gm \mu \sin\varepsilon \cos\varepsilon.
\end{equation} 

As previously outlined, up to the smallest order in $\frac{r}{a}$ the two torques are equal (and obviously in the opposite direction). This implies that the angular momentum of the system is conserved, as we expected.\\

Moreover, $\mu$ represents the mass of the ocean bulge, that is the mass of water whose shape is that of an ellipsoid (of semi-axis $R, R, R+h$) from which it is subtracted a sphere of radius $R$ concentric to it, with $h$ represents the tidal height of the ocean:
\begin{equation}
	\label{h}
	h=\frac{3}{2}\frac{m}{M} \biggl( \frac{R}{a} \biggr)^3 R.
\end{equation}
For a detailed computation of $h$ based on elementary mathematical tools see \cite{3}. 
So:
\begin{equation}
	\label{mu}
	\mu=\rho_w \frac{4}{3}\pi R^2h=\rho_w \frac{4}{3}\pi R^2 \frac{3}{2}\frac{m}{\rho_E \frac{4}{3}		\pi R^3} \biggl( \frac{R}{a} \biggr)^3 R = \frac{\rho_w}{\rho_E} \frac{3}{2} \biggl( \frac{R}{a} 	\biggr)^3 m.
\end{equation}

Finally, remembering that, according to our assumption, $r=R$ and using (\ref{mu}), the torque can be written as:
\begin{equation}
	\label{gamma}
	\Gamma=3\frac{R^2}{a^3}Gm \mu \sin\varepsilon \cos\varepsilon = \biggl( \frac{9}{2} \frac{\rho_w}{\rho_E}\biggr) G \frac{R^5}{a^6} m^2 \sin(2\varepsilon) = k G \frac{R^5}{a^6} m^2 \sin(2\varepsilon),
\end{equation}
where $k = \frac{9}{2} \frac{\rho_w}{\rho_E}$ is a dimensionless constant.\\
The formula (\ref{gamma}) is known in literature as "MacDonald formula for body-tide torques".
We note that our dumbbell model allows us to derive this formula in a simple way starting from reasonable physical considerations.\\

\section{Evolution of the system}
\label{section_evolution}
\begin{figure}
	\centering
	\includegraphics[scale=0.5]{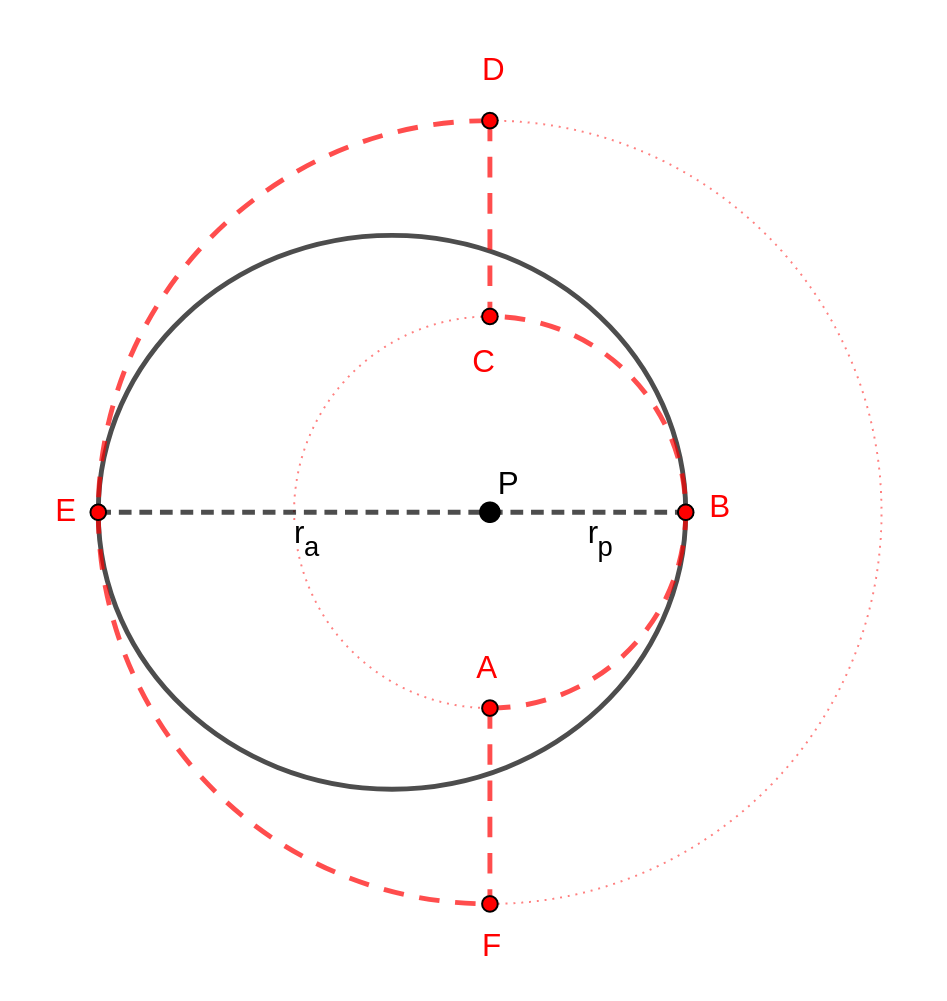}
	\caption{The real trajectory of the Moon around the Earth is the ellipse in black, the point $P$ represents the focus occupied by the Earth. The virtual trajectory is the dashed line in red, composed by the semi-circumference $ABC$ centered in $P$ and tangent to the ellipse in the perigee $B$ and the semi-circumference $DEF$ centered in $P$ and tangent to the ellipse in the apogee $E$.}
	\label{fig_trajectory}
\end{figure}

In this section we want to study the evolution of the system avoiding advanced mathematical tools. In order to do this, we imagine that the variations of parameters is not continuous but discrete in time. This can be done because the parameters vary on very large time scales, therefore at each revolution they vary by very small quantities. For this reason the difference between a discrete and a continuous evolution is irrelevant.\\
We start proving by this attitude the circularization of the orbit.\\
Since the results of the previous section hold in the case of circular orbit, we can imagine that the real elliptical orbit of the Moon is the superposition of two virtual semicircular orbits centered on the Earth and tangent to the real trajectory in the perigee and in the apogee respectively, as shown in figure \ref{fig_trajectory}. In this way we can apply the results obtained in the previous section, but unfortunately we have introduced a discontinuity in the trajectory of the Moon, which is very difficult to digest. However, imagining that the evolution of the parameters occur in a discrete way at each semi-revolution, the discontinuity in the virtual trajectory is irrelevant.\\

So: $r_a$ represent the Earth-Moon distance in the apogee and $r_p$ represent the Earth-Moon distance in the perigee. 
Clearly: $r_a - r_p > 0$.\\
Moreover:\\
$a = \frac{1}{2} (r_a + r_p)$ represent the semi-major axis of the orbit;\\
$c = \frac{1}{2}|F_1-F_2| = \frac{1}{2}(r_a - r_p)$ represent the half focal distance;\\
$e = \frac{c}{a}$ represent the eccentricity of the orbit.\\
To determine the evolution of the system, the torque plays a crucial role. As we calculated in the previous section, the torque acting on the Moon in the perigee is:
\begin{equation} \label{eq.moontorque}
    \Gamma_{M_p} = k G \frac{R^5}{r_p^6} m^2 \sin(2\varepsilon).
\end{equation}
At the same time, the torque acting on the dumbbell is: 
\begin{equation} \label{eq.dumbbelltorque}
    \Gamma_{D_p} = -k G \frac{R^5}{r_p^6} m^2 \sin(2\varepsilon).
\end{equation}
obviously the signs $\pm$ are arbitrary: in any case the torques have opposite sign.\\
The torques in the apogee ($\Gamma_{M_a}$ and $\Gamma_{D_a}$) are similar: just replace $r_p$ with $r_a$.\\
$\Gamma_M$ determines the variation of the orbital parameters (semi-major axis, focal distance, eccentricity); $\Gamma_D$ determines the variation of $\Omega$, the sidereal angular velocity of the Earth.\\
To determine the evolution of the parameters, we compare their values 
during the $n$-th revolution ($a_n; c_n; e_n; \Omega_n$) with their values during the $(n+1)$-th revolution ($a_{n+1}; c_{n+1}; e_{n+1}; \Omega_{n+1}$).\\
As we argued before, we imagine that the change of the parameters at the end of $n$-th revolution consist of two contribution at the end of each virtual semi-circumference.\\ 
Moreover: $r_{a,n}$ represents the Earth-Moon distance at the apogee during the $n$-th revolution;\\
$r_{p,n}$ represents the Earth-Moon distance at the perigee during the $n$-th revolution;\\
$a_n=\frac{1}{2}(r_{a,n}+r_{p,n})$ represents the semi-major axis during the $n$-th revolution;\\
$c_n=r_{a,n}-r_{p,n}$ represents the half focal distance during the $n$-th revolution;\\
$e_n=\frac{c_n}{a_n}$ represents the eccentricity during the $n$-th revolution.\\
The same parameters during the $(n+1)$-th revolution are indicated with the same notation, replacing the subscript $n$ with $(n+1)$.\\
Finally, the change of parameters between two successive revolutions are indicated with $\Delta$:\\
$\Delta r_p = r_{p,n+1} - r_{p,n}$; $\Delta r_a = r_{a,n+1} - r_{a,n}$; $\Delta a = a_{n+1} - a_n$; $\Delta c = c_{n+1} - c_n$; $\Delta e = e_{n+1} - e_n$.\\
We can start by considering the evolution of lunar orbital parameters. From now on, we indicate generically with $r_i$ the Earth-Moon distance in the position $i$: $i$ can be either $p$ (the perigee) or $a$ (the apogee).\\
Let $\Gamma_{Mn} = k G \frac{R^5}{r_i^6} m^2 \sin(2\varepsilon)$ the torque acting on the Moon in the $n$-th revolution. Then the variation of Moon's angular momentum
$L=m \omega r_i^2$ between the $n$-th and the $(n+1)$-th revolution is $\Delta L = L_{n+1} - L_n$ is:
\begin{equation}\label{eq.angularmomentum}
    \Delta L = \Gamma_{Mn}  T = \Gamma_{Mn} \frac{2\pi}{\omega}. 
\end{equation}
From this equation we can derive the evolution of $a$, the semi-major axis of the orbit. We remember that $\omega$ depends on $r_i$, indeed from the Kepler's third laws $\omega^2 r_i^3 = GM$. \footnote{This result holds in the case of a circular orbit. Thanks to the assumptions made above, this is our case.} Therefore we have:
\begin{equation}\label{eq.omega}
    \omega=\frac{\sqrt{GM}}{r_i^{3/2}}
\end{equation}
Hence, from \eqref{eq.angularmomentum} and \eqref{eq.omega}, we have:
\begin{equation}\label{eq.evolution_a}
    \Delta\left(m \sqrt{GMr_i}\right) = k G \frac{R^5}{r_i^6} m^2 \sin(2\varepsilon)\frac{2\pi r_i^{3/2}}{\sqrt{GM}},
\end{equation}
that we can rewrite as:
\begin{equation}\label{eq.evolution_a1}
    \Delta\left(\sqrt{r_i}\right) = 2\pi k \frac{m}{M} \frac{R^5}{r_i^{9/2}} \sin(2\varepsilon).
\end{equation}
We can rewrite the l.h.s. of the previous equation as:
\begin{align}
    \label{eq.derivative}
    \begin{aligned}
        &\Delta\left(\sqrt{r_i}\right) = 
        \sqrt{r_{i, n+1}}-\sqrt{r_{i,n}} = \sqrt{r_{i,n}+\Delta r_i}-\sqrt{r_{i,n}} = \sqrt{r_{i,n}} \left( \sqrt{1+ \frac{\Delta r_i}{r_{i,n}}} - 1 \right) \\
        &\simeq \sqrt{r_{i,n}} \left( 1+ \frac{1}{2}\frac{\Delta r_i}{r_{i,n}} - 1 \right) = \frac{\Delta r_i}{2\sqrt{r_{i,n}}} = \frac{\Delta r_i}{2\sqrt{r_i}},
    \end{aligned}
\end{align}
where we have used the approximation: $\sqrt{1+x}\simeq 1+\frac{1}{2}x$ (see appendix), with $x=\frac{\Delta r_i}{r_{i,n}} << 1$.\\
Hence equation \eqref{eq.evolution_a1} becomes:
\begin{equation}\label{eq.evolution_a2}
    \Delta r_i = 4\pi k \frac{m}{M} \frac{R^5}{r_i^4} \sin(2\varepsilon) = \frac{K}{r_i^4},
\end{equation} 
with $K= 4\pi k \frac{m}{M} R^5 \sin(2\varepsilon) > 0$ constant independent of $r_i$.\\
Actually $K$ depends on $\varepsilon$ which depends on $r_i$, indeed $\varepsilon$ is the difference between the angular position on the Moon (with respect a certain reference axis), $ \omega t $, and the sidereal angular position of the Earth, $\Omega t$:
\begin{equation}\label{eq_approxepsilon}
    \varepsilon = \Omega t - \omega t = \Omega t ( 1 - \frac{\omega}{\Omega}).  
\end{equation}
But if we suppose that $\omega << \Omega$ (this assumption is currently true for the Earth-Moon system, indeed: $\frac{\omega}{\Omega} \simeq \frac{1\,\text{day}}{1\,\text{month}} \simeq \frac{1}{30} \simeq 0,03$), then $\varepsilon \simeq \Omega t$ is independent on $r_i$.\\
Let us now suppose, as we argued before, that the variation of $a$ consists of two contributions: the variation when the Moon is at perigee ($\Delta a_p$) and the variation when the Moon is at apogee ($\Delta a_a$):
\begin{equation}\label{eq.variation_adouble}
\begin{cases}
    \Delta r_p = \frac{K}{r_p^4}\\
    \Delta r_a = \frac{K}{r_a^4}
\end{cases}
\end{equation}
But $r_a>r_p$, then: 
\begin{equation}\label{eq.deltaa}
    \Delta r_p>\Delta r_a.
\end{equation}
From equation \eqref{eq.deltaa} we can derive two important results:\\
First, $\Delta a = \Delta r_a + \Delta r_p > 0 $, this implies
\begin{equation}\label{eq.aincreases}
    a_{n+1} > a_n,
\end{equation} 
then the semi-major axis of the orbit increases.\\
Second, $\Delta c = c_{n+1} - c_n = \frac{1}{2}\left(r_{a,n+1} - r_{p,n+1}\right) - \frac{1}{2}\left(r_{a,n} - r_{p,n} \right) = \frac{1}{2}\left(r_{a,n+1} - r_{a,n}\right) - \frac{1}{2}\left(r_{p,n+1} - r_{p,n} \right) = \frac{1}{2}(\Delta r_a - \Delta r_p) < 0$, this implies
\begin{equation}\label{eq.cdecreases}
    c_{n+1} < c_n,
\end{equation}  
then the focal distance of the orbit decreases.\\
Finally, the two previous results implies that the eccentricity decreases and so the orbit becomes circular. Indeed:\\
\begin{align*}
    \Delta e 
    &= e_{n+1} - e_n = \frac{c_{n+1}}{a_{n+1}} - \frac{c_n}{a_n} = \frac{c_{n+1}a_n - c_na_{n+1}}{a_{n+1}a_n} = \frac{c_{n+1}a_n - c_na_n + c_na_n - c_na_{n+1}}{a_{n+1}a_n}\\
    &= \frac{a_n(c_{n+1} - c_n) - c_n(a_{n+1} - a_n)}{a_{n+1}a_n} = \frac{a_n \Delta c_n - c_n \Delta a_n}{a_{n+1}a_n} < 0,
\end{align*}
the last inequality follow from the fact that all the terms are positive, except for $\Delta c$. This implies
\begin{equation}\label{eq.edecreases}
    e_{n+1} < e_n,
\end{equation}  
then the eccentricity of the orbit decreases.\\
Moreover, we can determine the rate of decrease of $c_n$. Indeed:
\begin{align*}\label{eq.decreasingofc}
    \Delta c 
    &= \Delta (r_a - r_p) = \Delta r_a - \Delta r_p = \frac{K}{r_a^4} - \frac{K}{r_p^4} = \frac{K (r_p^4 - r_a^4)}{r_a^4r_p^4}\\
    &= \frac{K (r_p^2 + r_a^2)(r_p + r_a)(r_p - r_a)}{r_a^4r_p^4} = -c\frac{Ka(r_p^2 + r_a^2)}{r_a^4r_p^4} \simeq -c\frac{K}{a^5}= -\lambda c,
\end{align*}
with $\lambda = \frac{K}{a^5} > 0$.\\
Therefore $\Delta c_n$ decreases in a way directly proportional to $c_n$: this kind of decrease is called "exponential decrease
" where $\lambda$ is the rate of decrease. Since $\lambda$ in positive, then $c_n$ decreases until it becomes lesser than any prefixed positive quantity. When $c_n$ approaches $0$, then $r_{a,n} \simeq r_{p,n}$ and hence, from \eqref{eq.variation_adouble}, $\Delta r_a \simeq \Delta r_p$. This implies $\Delta c \simeq 0$. Thus, when the focal distance "becomes" zero, it no longer varies.\\
But $c_n \simeq 0$, $\Delta c \simeq 0$ imply that $e_n \simeq 0$, $\Delta e \simeq 0$. Thus even the eccentricity of the orbit decreases until it becomes lesser than any prefixed positive quantity and, "at the end" it no longer varies. So the orbit becomes circular.\\
On the other hand, the semi-major axis $a$ increases indefinitely: even when the orbit becomes circular, $a$ (that is its radius) continues to increase.\\
But the growth of $a$ and the decrease of $e$ occur on different time scales and in different ways, indeed while the variation of $e$ is exponential, i.e. very fast, that of $a$ is polynomial
\footnote
{
Indeed,
\begin{equation*}
    \Delta a = \frac{1}{2} \Delta (r_a + r_p) = \frac{1}{2} ( \Delta r_a + \Delta r_p ) = \frac{1}{2} \left( \frac{K}{r_a^4} + \frac{K}{r_p^4} \right) = \frac{K (r_p^4 + r_a^4)}{2r_a^4r_p^4} \simeq \frac{K}{2a^4}.
\end{equation*}
},
which is slower.\\
Let us now study the evolution of $\Omega$, the sidereal angular velocity of the Earth: it's variation is due to $\Gamma_D$, the torque acting on the dumbbell. Indeed the dumbbell is pulled back from the Moon and thanks to the friction between it and the underlying planet, it slows down the rotation of the Earth.\\
The angular momentum of the Earth is $I_E \Omega$, where $I_E$ is the moment of inertia. As we argued before, we suppose that it varies in each lunar revolution. Thus, at the end of $n$-th revolution we have:
\begin{equation}\label{eq.angularspeed1}
    \Delta \left( I_E \Omega \right) = \Gamma_{D,n}  T = \Gamma_{D,n} \frac{2\pi}{\omega},
\end{equation}
and hence:
\begin{equation}\label{eq.angularspeed2}
    \Delta \Omega = \frac{\Gamma_{D,n}}{I_E}\frac{2\pi}{\omega}=-2\pi k G \frac{R^5}{r_i^{9/2}} \frac{m^2}{I_E} \sin(2\varepsilon) < 0.
\end{equation}
So, as we expect, $\Omega$ decrease.\\
But if $\Omega$ decreases, also $\varepsilon$ decreases; indeed as argued before (see equation \eqref{eq_approxepsilon}), $\varepsilon \simeq \Omega t$, then:
\begin{equation}
    \Delta \varepsilon = \varepsilon_{n+1} - \varepsilon_n = \Omega_{n+1} t - \Omega_n t  = (\Omega_{n+1} - \Omega_n)t = \Delta \Omega t < 0, 
\end{equation}
this implies
\begin{equation}\label{eq.epsilondecreases}
    \varepsilon_{n+1} < \varepsilon_n,
\end{equation}  
then $\varepsilon$ decreases.\\
This dynamics has as a stationary situation that in which $\varepsilon = 0$. Indeed, when $\varepsilon$ approaches $0$, then $\Delta \Omega$ becomes $0$ (see equation \eqref{eq.angularspeed2}), and then $\Omega$ no longer varies. But, from equation \eqref{eq_approxepsilon}, when $\varepsilon=0$ then $\Omega - \omega = 0$ that implies $\Omega = \omega$.\\
So the stationary situation is the spin-orbit resonance regime: that is a dynamics in which the Earth complete a rotation over a period of time equal to a Moon revolution. In such a situation the dumbbell (that rotates with velocity $\omega$) moves on the Earth at the same speed as that of the underlying layer (that rotates with velocity $\Omega$) and thus friction between the dumbbell and the underlying Earth ends: in such a situation tides will no longer be observed on Earth and no more energy will be dissipated by this mechanism.\\ 
This type of dynamics is often observed in the oldest planet-satellite systems of the Solar System, indicating the fact that the equilibrium situation predicted by our model is indeed attractive.\\

\appendix

\section{Special approximation}
In this section we want to show two ways to prove the special approximation we used in equation \eqref{eq.derivative}: 
\begin{equation}\label{eq.approx}
    \sqrt{1+x} \simeq 1 + \frac{1}{2}x,
\end{equation}
valid if $x<<1$.\\
The first way uses the linearization formula: given a function $f(x)$,  if in its point $x_0$ the function $f$ and its derivative $f'$ are both continuous, it is possible to approximate the function with the tangent line in $x_0$:
\begin{equation}\label{eq.tangent}
    f(x_0+\delta) = f(x_0) + f'(x_0)\delta + R(\delta),
\end{equation}
with $R(\delta) = o(\delta^2)$.\\
So equation \eqref{eq.approx} follows from \eqref{eq.tangent} if $f(x) = \sqrt{1+x}$, $x_0 = 0$ and $\delta = x << 1$.\\
A second way to prove \eqref{eq.approx} does not use derivative, but only techniques of Cartesian geometry. The issue consists to determine the tangent line to the function $y=\sqrt{1+x}$ in its point $x_0=0$.\\
In the Cartesian plane, $y=\sqrt{1+x}$ represents the equation of a semi-parabola. The tangent point is $(0, 1)$ and then the equation of tangent line is: $y=mx+1$. $m$ can be determined by imposing the tangent condition.\\
We can rewrite the equation of our semi-parabola as $y^2=1+x$ and we can substitute $y$ with its expression $mx+1$, obtaining:
\begin{equation}\label{eq.resolv}
    m^2x^2+(2m-1)x=0.
\end{equation}
Finally we have to impose the tangent condition (uniqueness of the solution): the discriminant of \eqref{eq.resolv} is null. This implies: $m=1/2$.\\
Then, the equation of tangent line is $y=\frac{1}{2}x+1$.\\
So, the semi-parabola $y=\sqrt{1+x}$ can be approximated near the point $x_0=0$ with its tangent line $y=\frac{1}{2}x+1$.\\

The authors have no conflicts to disclose.

\end{document}